\documentclass[aps,pre,twocolumn,groupedaddress,showpacs,showkeys]{revtex4}
\usepackage{graphicx}
\usepackage{color}
\usepackage{amssymb,amsfonts,amsmath}

\begin{document}

\title{Tolerating the Community Detection Resolution Limit with
	Edge Weighting}

\author{Jonathan W. Berry}\email{jberry@sandia.gov}
\affiliation{Sandia National Laboratories, P.O. Box 5800, Albuquerque, NM, 87185.}
\thanks{This manuscript has been authored by Sandia Corporation under Contract No. DE-AC04-94AL85000 with the U.S. Department of Energy.  The United States Government retains and the publisher, by accepting the article for publication, acknowledges that the United States Government retains a non-exclusive, paid-up, irrevocable, world-wide license to publish or reproduce the published form of this manuscript, or allow others to do so, for United States Government purposes.}
\author{Bruce Hendrickson}
\author{Randall A. LaViolette}
\author{Cynthia A. Phillips}
\affiliation{Sandia National Laboratories, P.O. Box 5800, Albuquerque, NM, 87185.}
\date{\today}
\begin{abstract} 

Communities of vertices within a giant network such as the 
World-Wide Web are likely to be vastly smaller than the network itself.
However, Fortunato and Barth\'{e}lemy have proved that modularity maximization
algorithms for community detection
may fail to resolve communities with
fewer than $\sqrt{L/2}$ edges, where $L$ is the number of edges
in the entire network.  This resolution limit leads modularity maximization 
algorithms to have notoriously poor accuracy on many real networks.
Fortunato and Barth\'{e}lemy's argument can be extended to networks with
weighted edges as well, and we derive this corollary argument. We conclude
that weighted modularity algorithms may fail to resolve communities
with fewer than $\sqrt{W \epsilon/2}$ total edge weight, where $W$
is the total edge weight in the network and $\epsilon$ is the maximum weight
of an inter-community edge. If $\epsilon$ is small, then small communities
can be resolved.

Given a weighted or unweighted network, we describe how to derive new edge 
weights in
order to achieve a low $\epsilon$, we modify the ``CNM'' community detection
algorithm to maximize weighted modularity, and show that the resulting
algorithm has greatly improved accuracy.  In experiments with an emerging
community standard benchmark, we find that our simple CNM variant is 
competitive with the most accurate community detection methods yet proposed. 

\end{abstract}

\maketitle

\section{Introduction}
\label{sec:intro}

Maximizing the modularity of a network, as defined by Girvan and 
Newman~\cite{ng2004}, is perhaps the most popular and cited 
paradigm for detecting communities in networks.  There are many algorithms
for approximately maximizing modularity and its variants, 
such as~\cite{cnm2004,c2005,flzwd2006}.
Community assignments of good modularity
feature groups of nodes that are more tightly connected than would be expected.
We give the formal definition of modularity below.
Recent literature, however, 
has begun to focus on paradigms other than modularity maximization.  This
is in part due to Clauset, Newman, and Moore~\cite{cmn-nature}, who now 
advocate a more general notion of ``community'' than that 
associated with modularity. The 
shift away from modularity maximization is also due to
Fortunato and Barth\`{e}lemy~\cite{fb2007}, who prove
that any community assignment produced by a modularity maximization algorithm 
will have predictable deficiencies in certain realistic 
situations. Specifically, they argue that any solution of maximum modularity
will suffer from a {\em resolution limit} that prevents small communities
from being detected in large networks.  Furthermore, 
work by Dunbar~\cite{d1998} indicates that true human communities are 
generally smaller that 150 nodes.  This size is far below the resolution
limit inherent in many large networks, such as various social 
networking sites on the World Wide Web.

We agree with Clauset, Newman, and Moore's~\cite{cmn-nature} idea that it is useful to
consider more general definitions for ``community''; however, we maintain
that it is still important to detect traditional, tightly-connected
communities of nodes.  In this paper, we revisit the negative result of 
Fortunato and Barth\`{e}lemy and analyze it in a different light.  
We show that positive results are 
possible without contradicting the resolution limit.  
The key is to
apply carefully computed weights to the edges of the network.

With one exception, previous methods for tolerating this resolution limit 
require searching over an input parameter.
For example, Li, et al.~\cite{lzwzc2008} address the 
resolution limit problem by defining a modularity alternative called 
{\em modularity density}. Given a fixed number of communities $k$,
solving a k-means problem will maximize modularity density.  Li, et al.
generalize modularity density so that tuning a parameter $\lambda$
favors either small communities (large $\lambda$) or large communities
(small $\lambda$)~\cite{lzwzc2008}.
Arenas, Fernandez, and Gomez also address the problem of resolution 
limits~\cite{afg2008}.  They provide the user with a parameter $r$
that modifies the natural community sizes for modularity maximization 
algorithms.  By tuning $r$, they influence the natural 
resolution limit.  
At certain values of $r$, small communities will be
natural, and at other values of $r$, large communities will be natural.
Our methods apply without specifying any target scale for natural
communities, and resolve small and large communities simultaneously.

One solution that resolves communities at multiple scales
with no tuning parameter is the HQcut algorithm of Ruan and Zhang~\cite{rz2008}.
This algorithm alternates between spectral methods and
efficient local improvement.  It 
uses a statistical test to determine whether to split each
community.  Ruan and Zhang argue that a subnetwork with modularity 
significantly greater than that expected of a random network with the same 
sequence of vertex degrees is likely to have sub-communities, and therefore 
should be split.  As Fortunato points out in his recent survey~\cite{f2009}, 
though, this stopping criterion is an ad-hoc construction.  

Nevertheless, Ruan and Zhang present compelling evidence that the accuracy of 
HQcut often exceeds that of competitors such as Newman's spectral method 
followed by Kernighan-Lin local improvement~\cite{n2006} and the simulated 
annealing method of Guimer{\`a} and Amaral~\cite{ga2005}.  The HQcut solution
is not simply the solution of global maximum modularity, so it is not
bound by the resolution limit.  We obtained the authors' Matlab code for
HQcut and we present comparisons with our approach below.

\section{Resolution Limits}
\label{sec:resolution}

Fortunato and Barth\'{e}lemy~\cite{fb2007} define a {\em module} to be a set of vertices
with positive modularity:
\begin{equation}
\frac{l_s}{L} - \left(\frac{d_s}{2L}\right)^2 > 0,
\label{eq:module}
\end{equation}
where $l_s$ is the number of undirected edges (links) within the set,
$d_s$ is the sum of the degrees of the vertices within the set, and $L$
is the number of undirected links in the entire network.
These modules contain
more edges than we would expect from a set of vertices with the same degrees,
were edges to be assigned randomly (respecting the invariant vertex degrees).
Let us define such modules to be {\em natural communities} with respect to 
modularity maximization. We say that a natural community is {\em minimal}
if it contains no other natural communities. We wish to resolve the
minimal natural communities, and we will discuss this goal in 
Section~\ref{sec:lfr}.

In order to ensure that such modules
are resolved in a global community assigment with maximum modularity, Fortunato and Barth\'{e}lemy~\cite{fb2007} argue that the following must hold:
\begin{equation}
l_s \ge \sqrt{\frac{L}{2}}.
\label{eg:fb-limit}
\end{equation}
They back up this mathematical argument with empirical evidence.  
Even in a pathologically easy situation, in which the modules are cliques,
and only one edge links any module to a neighboring module, the individual
modules will not be resolved in any solution of maximum modularity.  Instead,
several cliques will be merged into one module. Experiments show that the 
numbers of links in the resulting modules closely track 
the $\sqrt{L/2}$ prediction.

Work by Dunbar~\cite{d1998} indicates that true human communities are generally
limited to roughly 150 members, and this is corroborated by the recent work
of Leskovec, Lang, Dasgupta, and Mahoney~\cite{lldm2008}.  Such communities
will have dramatically fewer than $\sqrt{L/2}$ edges in practice.
Based on this argument, it would seem that there is little hope for 
the solutions of modularity maximizing algorithms to be applied in real 
situations in which
$L \gg l_s$.  Indeed, partially due to the resolution limit result, the 
general direction of research in community detection
seems to have shifted away from modularity maximization in favor of 
machine learning techniques.

In this paper, we revisit the resolution limit in the context of edge 
weighting and derive more positive results.

\section{Resolution with edge weights}

The definition of a module in equation [\ref{eq:module}] can easily be 
generalized when
edges have weights.  Let $w_s$ be the sum of the weights of all undirected 
edges connecting vertices within Set $s$.  Let $d^w(v)$, the weighted
degree of vertex $v$, be the sum of the weights of all edges incident on $v$.
We define $d^w_s = \sum_{v \in s} d^w(v)$ to be the sum of weighted degrees
of the vertices in Set $s$.  Then Set $s$ is a module if and only if:
\begin{equation}
\frac{w_s}{W} - \left(\frac{d^w_s}{2W}\right)^2 > 0.
\label{eq:module2}
\end{equation}

Following~\cite{fb2007} step-by-step, when considering a module, 
we use $w^{\mbox{\scriptsize{out}}}_s$ 
to denote the sum of the weights of the edges leaving Set $s$, and 
also note that $w^{\mbox{\scriptsize{out}}}_s = \alpha_s w_s$, where $\alpha_s$ 
is a convenience that enables
us to rewrite the definition of a module in a useful way. We now have
$d^w_s = 2w_s + w^{\mbox{\scriptsize{out}}}_s = (\alpha_s + 2) w_s$, and a new, 
equivalent, definition of a module:
\begin{equation}
\frac{w_s}{W} - \left(\frac{(\alpha_s + 2) w_s}{2W}\right)^2 > 0.
\label{eq:module3}
\end{equation}

Manipulating the inequality, we obtain the relationship:

\begin{equation}
w_s < \frac{4W}{(\alpha_s + 2)^2}.
\label{eq:comm-bound}
\end{equation}

Thus, sets representing communities must not have too much weight in 
order to be modules.  

\section{The Maximum Weighted Modularity}
\label{sec:max-weight-modularity}

Fortunato and Barth\'{e}lemy describe the most modular network
possible.  This yields both computed figures that can be corroborated by 
experimental evidence, and intuition that the resolution limit in community
detection has a natural scale that is related to the total number of links 
in the network.  We will use the same strategy for the weighted case.

First, we imagine a network in which every module is a clique.  For a given 
number of nodes and number of cliques, the 
modularity will be maximized if each clique has the same size.  Weighting
does not change the argument of~\cite{fb2007} that the modularity
approaches 1.0 as the number of cliques goes to infinity.  Now, 
following~\cite{fb2007}, we consider a slight relaxation of the simple 
case above: the most modular connected network.  This will be our 
set of $m$ cliques
with at least $m-1$ edges to connect them.  Without loss of generality, 
we consider the case of $m$
connecting edges --- a ring of cliques, as studied by~\cite{ddda2005}.

Departing for a moment from~\cite{fb2007}, we now consider an 
edge weighting for the network.  With edge weights in the range 
$[0,1]$, the optimal weighting would assign 1 to each
intra-clique edge and 0 to each connecting edge.  The weighted 
modularity of this weighted network would be equivalent to the unweighted
modularity of the $m$ independent cliques described above, and would tend
to 1.

Relaxing this idealized condition, now assume that we have a weighting
function that assigns $\epsilon$ to each connecting edge, and 1.0 to each
intra-clique edge.  We now analyze the resulting weighted modularity.

The total edge weight contained within the cliques is

\begin{equation}
\sum_{s=1}^m{w_s} = {W} - \epsilon m.
\label{eq:total-weight-within}
\end{equation}

Each clique is a module by (\ref{eq:module2}) provided that $\epsilon$ is sufficiently small.
Summing the contributions of the 
modules, we find the weighted modularity of the network when broken into these
cliques is:

\begin{equation}
Q = \sum_s \left[\frac{w_s}{W} - \left(\frac{2 w_s + 2 \epsilon}{2W}\right)^2
			\right].
\label{eq:Q}
\end{equation}

Since all modules contain the same weight, for all $s$:

\begin{equation}
w_s = \frac{W - \epsilon m}{m} = \frac{W}{m} - \epsilon
\label{eq:each module}
\end{equation}

The maximum modularity of any solution with $m$ communities is:

\begin{equation}
Q_M(m,W) = m \left[\frac{W/m - \epsilon}{W} - 
	\left(\frac{W/m}{W}\right)^2\right] = 
		1 - \frac{\epsilon m}{W} - \frac{1}{m}
\label{eq:Q_m}
\end{equation}

To quantify this maximum, we take the derivative with respect to $m$:
\begin{equation}
\frac{dQ_M}{dm}(m,W) = \frac{-\epsilon}{W} + \frac{1}{m^2}
\label{eq:dQ_m}
\end{equation}

Setting this to zero, we find the number of communities in the 
optimal solution:
\begin{equation}
m^* = \sqrt{\frac{W}{\epsilon}}.
\label{eq:opt-num-communities}
\end{equation}

Substituting into (\ref{eq:Q_m}), we find the maximum 
possible weighted
modularity:

\begin{equation}
Q_M(W) = 1 - \frac{2}{\sqrt{W/\epsilon}}.
\label{eq:Q_M}
\end{equation}

The unweighted versions of equations~[\ref{eq:opt-num-communities}] and~[\ref{eq:Q_m}]
from ~\cite{fb2007} are, respectively, $m^* = \sqrt{L}$, and 
$Q_M(L) = 1 - \frac{2}{\sqrt{L}}$. In this unweighted case, the natural
scale is clearly related to $L$.  We don't expect to 
be able to find many more than $\sqrt{L}$ modules in any solution
of optimal unweighted modularity.

Our weighted case is similar, but the introduction of $\epsilon$
leads to some intriguing possibilities.  If $\epsilon$ can be made
small enough, for example, then there is no longer any limit to the
number of modules we might expect in any solution of maximum 
weighted modularity.

\section{The Weighted Resolution Limit}

In~\cite{fb2007}, Fortunato and Barth\'{e}lemy prove that 
any module in which $l < \sqrt{L/2}$ may not be resolved by 
algorithms that maximize modularity.  Their argument characterizes the
condition under which two true modules linked to each other by any 
positive number of edges will contribute more to the global modularity 
as one unit rather than as two separate units. 
This result is corroborated by experiment.
In a large real-world dataset such as the WWW, 
modules with $l \ll L$ will almost certainly exist.  

Following the arguments of~\cite{fb2007} directly, while considering
edge weights, we now argue that any module $s$ in which 
\begin{equation}
w_s < \sqrt{\frac{W\epsilon}{2}} - \epsilon
\label{eq:resolution}
\end{equation}
\noindent
may not be resolved. 
Consider a scenario in which two small modules are
either merged or not.  Suppose that the first module has intra-module edges 
of net weight $w_1$, and the second has intra-module edges of net weight
$w_2$. 
We assume that inter-module edges between these two modules have 
weight $\epsilon$, explicitly write the expressions for weighted 
modularity in both cases, and find their difference.  
The weighted modularity of the solution in which these 
two modules are resolved exceeds that in which they are merged, provided:
\begin{equation}
w < \frac{2W\epsilon/w}{(\frac{\epsilon}{w} + \frac{\epsilon}{w} + 2)
                          (\frac{\epsilon}{w} + \frac{\epsilon}{w} + 2)}
\end{equation}
where $w$ could be either $w_1$ or $w_2$. Manipulation of this expression gives~(\ref{eq:resolution}).

Two challenges remain: finding a method to set edge weights
that achieve a small $\epsilon$, and adapting modularity maximization
algorithms to use weights.  The second challenge is partially addressed
by~\cite{nw2006} and~\cite{flzwd2006}, but we take a different approach.

\section{Edge Weighting}
\label{sec:edge-weighting}
There are myriad ways to identify local structure with local computations.
Several approaches to community detection, such as~\cite{c2005,mrc2005,lfk2008},
are based upon this idea.  We use local computations to derive new edge 
weights.  Our approach is to reward an edge for each short cycle connecting its 
endpoints.  These suggest strong interconnections.

\begin{figure}[tbh]
\centerline{\includegraphics[width=2in]{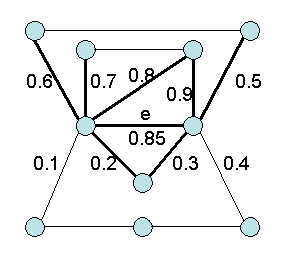}}
\caption{Edge neighborhood weighting}
\label{fig:neigh-cases}
\end{figure}

For a vertex $v$, let $E(v)$ be the 
set of all undirected edges incident on $v$.  We
also define the following sets to express
triangle and rectangle relationships between pairs
of edges.
\[T_e = \{e' : \mbox{there exists a 3-cycle containing both $e$ and $e'$}\} \]
\[R_e = \{e' : \mbox{there exists a 4-cycle containing both $e$ and $e'$}\} \]
Note that $e$ can be a member of $T_e$ and $R_e$.

The total weight of edges incident on the endpoints of edge $e = (u,v)$ is
\begin{eqnarray*}
	W_e & = & \sum_{e' \in E(u) \cup E(v)} w_{e'}.
\end{eqnarray*}
We consider incident edges that reside on paths of at most three edges 
connecting the 
endpoints of $e$ to be ``good'' with respect to $e$.
\begin{eqnarray*}
	G_e & = & \sum_{e' \in E(u) \cup E(v) \cap (T_e \cup R_e)} w_{e'}.
\end{eqnarray*}

Such edges add credence to the proposition that $e$ is an intra-community edge.
We define {\em neighborhood coherence} of $e$ as follows:
\[C(e) = \frac{G_e}{W_e}\].

For example, in
Figure ~\ref{fig:neigh-cases}, the coherence is computed by summing the
weights of the thickened edges and dividing by the total weight of edges
incident on the endpoints of $e$:
$C(e) = \frac{4.85}{5.35}.$
Alternate definitions are possible,
of course,  but this weighting is intuitive and performs well in practice.

Arenas, Fernandes, and Gomez,
by contrast, add self-loops to vertices according to their $r$ parameter, 
thereby ``weighting'' the nodes, and also adding more intra-community edges to
each module.  Thus, they pack more edges into each module in order to satisfy
Inequality~[\ref{eg:fb-limit}].

We have considered generalizing $C(e)$ to include cycles of length 5 and 
greater, but this would be a considerable computational expense, and
we expect diminishing marginal benefit.  

Now we give a simple iterative algorithm for computing
edge weights:
\label{sec:weighting}

\begin{enumerate}
\item Set $w_e = 1.0$ for each edge $e$ in the network (or accept $w_e$ as
	input if the edges are already weighted).
\item Compute $C(e)$ for each $e$, set $w_e = C(e)$.\label{loop:begin}
\item If any $w_e$'s changed within some tolerance, go to Step~\ref{loop:begin} \label{loop:end}
\end{enumerate}

This process will tend to siphon weight out of the inter-module edges (those
with smaller $C(e)$), thus lowering $\epsilon$. We find in practice that
it terminates in a small number of iterations.  Computing $C(e)$ reduces to
finding the triangles and 4-cycles in the graph.  This can be done naively
in $O(mn\log n)$ time on scale-free graphs.
We use Cohen's data structures~\cite{c2009} that admit more efficient
algorithms in practice.  For WWW-scale graphs, it may be necessary for
efficiency reasons to ignore edges incident on high-degree vertices.  
This would isolate these vertices. However, since such vertices 
often have special roles in real networks, they might require individual
attention anyway.

We define Algorithm $W(k)$ to be $k$ iterations through 
the loop in Steps~\ref{loop:begin}--\ref{loop:end}.

\section{Weighted Clauset-Newman-Moore}
\label{sec:wCNM}

Any modularity maximization algorithm could be made to 
leverage edge weights such as those computed in the previous section.  
Newman replaces individual weighted edges with sets of 
multiple edges, each with integral weight~\cite{nw2006}.  
We modify the agglomerative
algorithm of Clauset, Newman, and Moore (CNM)~\cite{cnm2004} to handle
arbitrary weights directly.

The CNM algorithm efficiently computes the change in modularity 
($\Delta Q$) associated with all possible mergers of two existing 
communities.  At the beginning, each vertex is in its own singleton 
community.  Unweighted modularity is defined as follows:
\begin{eqnarray*}
Q & = & \frac{1}{2L}\sum_{vw}\left[ A_{vw} - \frac{k_vk_w}{2L}\right]\delta(c_v,c_w) \\
  & = & \sum_s (e_{ss} - a^2_s).
\end{eqnarray*}
\noindent
$A_{vw}$ is the adjacency matrix entry for directed edge $(v,w)$, $k_v$
is the degree of vertex $v$, $e_{rs}$ is the 
fraction of edges that link vertices in community $r$ to vertices in community
$s$, and $a_s = \sum_r e_{rs}$ is the sum of the degrees of all vertices in 
community $s$ divided by the total degree. 
The function $\delta(c_v,c_w)$ equals $1$ 
if $v$ and $w$ are in the same community, and $0$ otherwise.

Since vertices $i$ and $j$ initially reside in their own singleton communities,
$e_{ij}$ is initially simply $\frac{A_{ij}}{2L}$.  The first step in CNM 
is to initialize $\Delta Q$ for all possible mergers: 
\begin{equation}
\Delta Q = \left\{ \begin{array}{l}
			1/(L) - 2k_ik_j/(2L)^2 \ \ \mbox{if $i$,$j$ are connected}\\
			0  \ \ \mbox{otherwise.}
			\end{array}\right.
\end{equation}

\begin{table}[tbh]
\centerline{
\begin{tabular}{|l|l|l|l|l|l|l|} \hline
Algorithm & $\epsilon$ &        $m^*$   &       $|S|$ & $Q_M$ & $Q$  \\ \hline
CNM       &  N.A. &   108 & 108 & 0.980 & 0.980  \\ \hline
wCNM$_1$ & 0.111 & 286 & 263 & 0.9930 & 0.9928 \\ \hline
wCNM$_5$ & $<0.000001$ & 1000 & 1000  & 0.9999 & 0.9986 \\ \hline
\end{tabular}
}

\caption{These results from the ring of 1000 5-cliques illustrate gains 
	made by considering weighting.  
	Predicted
	($m^*$) and algorithmically discovered ($|S|$) numbers of communities 
	match well and 
	indicate that careful weighting makes it possible to resolve all
	1000 cliques as modules in a solution of maximal weighted modularity.
	$Q_M$ is defined in (\ref{eq:Q_M}), $m^*$ is defined 
	in (\ref{eq:opt-num-communities}), and $\epsilon$ is the weight
	of the heaviest edge between two communities. 
	}
\label{tab:cliques}
\end{table}

CNM also initializes $a_i = \frac{k_i}{2L}$ for each vertex $i$. Once the 
initializations are complete, the algorithm repeatedly selects the best
merger, then updates the $\Delta Q$ and $a_i$ values, until only one 
community remains.  The solution is the community assigment with the 
largest value of $Q$ encountered during this process.  Clever data
structures allow efficient update of the $\Delta Q$ values.

To modify CNM to work on weighted graphs, we need only change
the initialization step.  The update steps are identical.  We simply define
and compute the weighted degree of each vertex $k^w_i = \sum_j w_{ij}$.  The
initialization becomes:
\begin{equation}
\Delta Q^w = \left\{ \begin{array}{l}
			w_{ij}/(W) - 2k^w_ik^w_j/(2W)^2 \ \ \mbox{if $i$,$j$ are connected}\\
			0  \ \ \mbox{otherwise,}
			\end{array}\right.
\end{equation}
and $a^w_i = \frac{k^w_i}{2W}$.  With these initializations, normal 
CNM merging greedily maximizes weighted modularity $Q^w$.  We
refer to this algorithm as wCNM.  Note that our definition of $Q^w$ is
equivalent to that of~\cite{flzwd2006}.

\section{Results}

Given an undirected, weighted or unweighted network, 
we apply the Algorithm $W(k)$ to set our edge weights, then run wCNM.
We use wCNM$_k$ to denote this two-step process.
Note that running wCNM$_0$ is equivalent to running CNM.

We will consider two different datasets:
the ring of cliques example discussed above, and
the benchmark of~\cite{lfr2008}, which is a generalization of the 128-node
benchmark of Girvan and Newman~\cite{gn2002}.

\subsection{The ring of cliques}
Refer to 
Table ~\ref{tab:cliques} 
for the following discussion.  
Danon, D\'{i}az-Guilera, Duch, and Arenas ~\cite{ddda2005} considered 
$m$ disconnected cliques as a pathological 
example of maximum modularity (which approaches 1.0 as the number of cliques
increases).  Fortunato and Barth\`{e}lemy~\cite{fb2007} add single 
connections between cliques to form a ring.
Our intuition is that the natural communities in such
a graph are the cliques.  However, the resolution limit argument of Fortunato
and Barth\`{e}lemy indicates that this will not be the solution of maximum
modularity if each clique has fewer than $\frac{\sqrt{L}}{2}$ edges.  
They confirm this via experiment, and we have reproduced their 
results for an instance with 1000 cliques of size five. Table~\ref{tab:cliques}
summarizes the performance of CNM and wCNM for this case.
The $m^*$ column
contains the number of communities expected in a solution of maximum weighted
modularity, as defined in~\ref{eq:opt-num-communities}.  The first row
shows the unweighted case, in which $m^*$ is equivalent to that defined
in~\cite{fb2007}.  CNM achieves this theoretical maximum by finding 108
communities, which is much smaller than the number of cliques.

If we run wCNM$_1$, which performs one iteration of neighborhood coherence, we
obtain the results in Row 2 of Table ~\ref{tab:cliques}. 
The value of 
$\epsilon$ we observe is 0.047, leading via~(\ref{eq:opt-num-communities})
to an estimate of 286 resolved 
communities.  The wCNM$_1$ algorithm resolves 263.  In a run with
five iterations, labeled 
wCNM$_5$, we
both expect and find 1000 communities, resolving all of the natural communities
and simultaneously observing our highest weighted modularity.  Iterating
further reduces $\epsilon$ without changing the community assignment.

\begin{figure}[tbh]
\centerline{\includegraphics[width=3.5in]{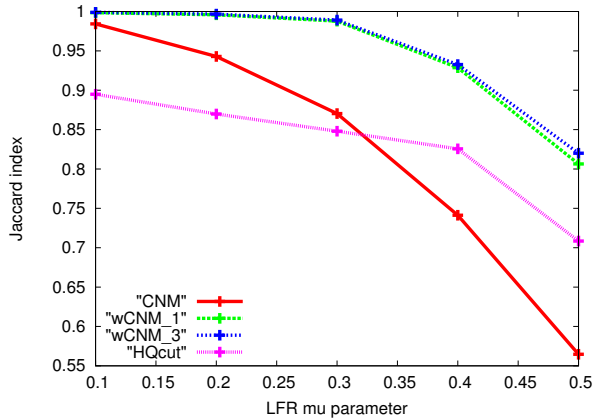}}
\caption{Mutual information study for the LFR benchmark.
\label{fig:mutual}}
\end{figure}

\begin{figure*}[tbh]
\centerline{
\begin{tabular}{ccc} \\
\includegraphics[width=2.25in]{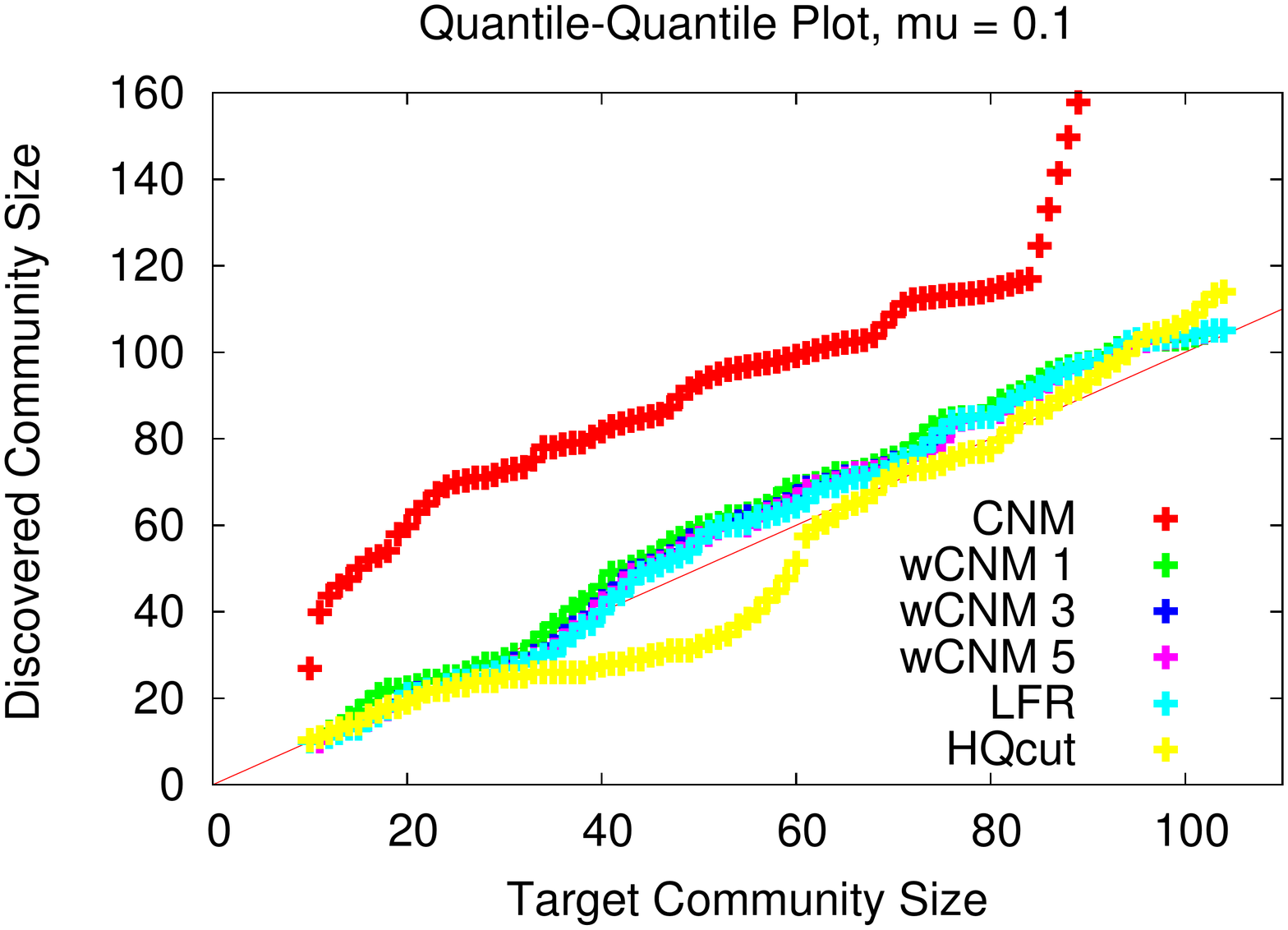} &
\includegraphics[width=2.25in]{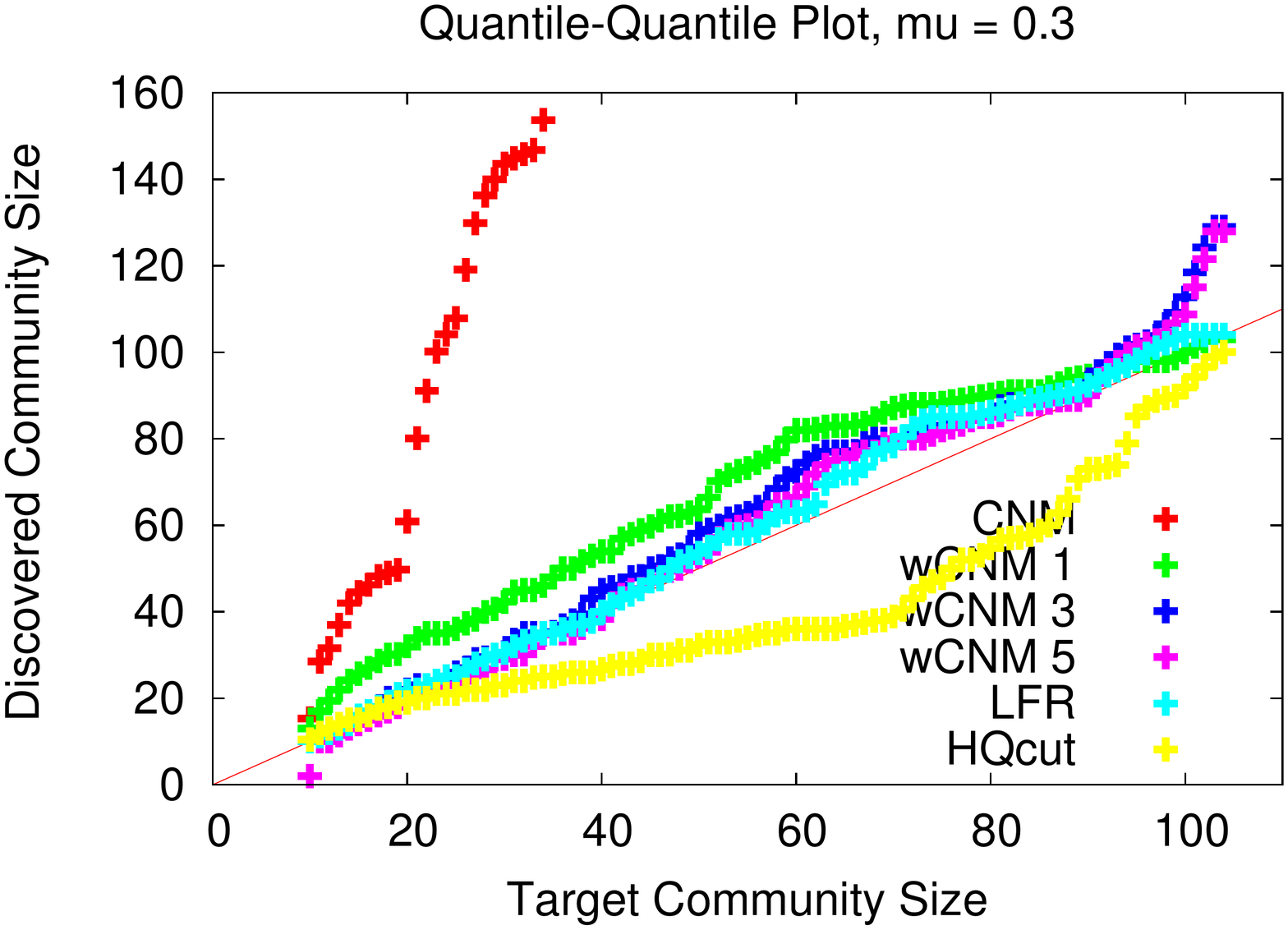} &
\includegraphics[width=2.25in]{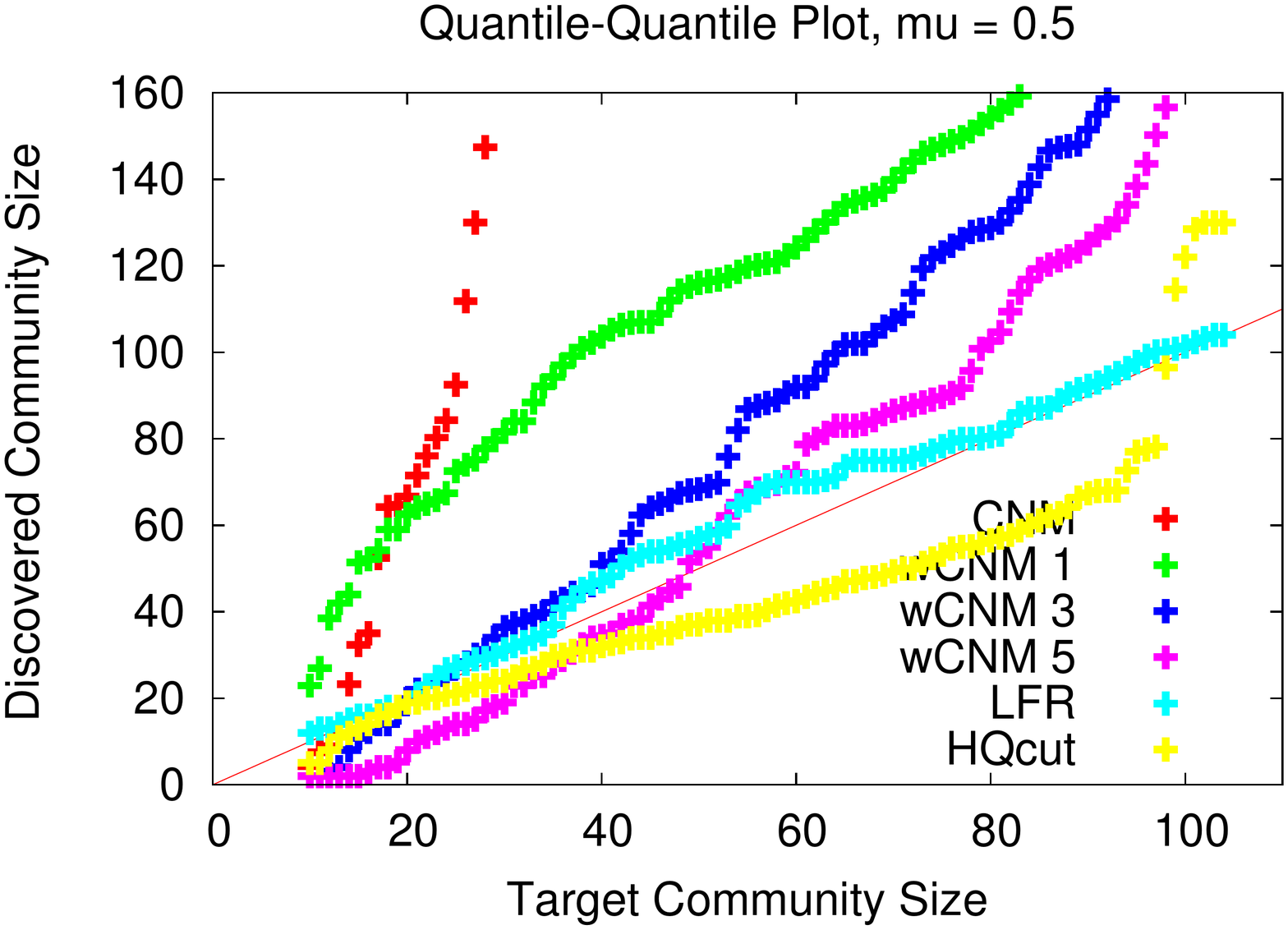}  \\
(a) & (b) & (c)
\end{tabular}
}
\caption{Example distributions of community sizes are shown in these 
quantile-quantile plots.
The line $y=x$ represents a perfect match between discovered community
sizes and the LFR power-law distribution.
\label{fig:cdf}}
\end{figure*}

\subsection{The LFR Benchmark}
\label{sec:lfr}
Lancichinetti, Fortunato, and Radicchi~\cite{lfr2008} (LFR) give a 
generalization of the popular Girvan and Newman benchmark~\cite{gn2002} 
for evaluating community detection algorithms.  The latter consists of
128-vertex random graphs, each with 4 natural communities of size 32.  The
user tunes a parameter to adjust the numbers of intra-community and 
inter-community edges.  
Many authors use this benchmark to create 
plots of ``mutual information,'' or agreement in node classification between 
algorithm-discovered communities and natural communities.  The LFR
benchmark is similar in spirit, but considerably more realistic.  It
allows the user to specify distributions both for the community sizes
and the vertex degrees. Users also specify 
the average ratio (per vertex) of 
inter-community adjacencies to total adjacencies, called 
{\em mixing parameter} $\mu$.  
At $\mu = 0.0$, all edges are intra-community.

The LFR benchmark construction process begins by sampling vertex degrees
and creating a graph with the selected degree distribution.
It then samples community sizes.
A vertex of degree $k$ should have about $(1-\mu)k$ neighbors
from the same community.  Therefore, it is assigned to a 
community with at least $(1-\mu)k+1$ vertices. 
LFR assigns vertices to communities via an interated random
process enforcing this constraint, then rewires
until the average $\mu$ meets the desired value.
We have a special interest in the LFR benchmark because it generates graphs
with both small and large natural communities.  

For several different values of $\mu$, we used the C code from Fortunato's 
web site (cited in~\cite{lfr2008}) to generate 30 instances 
each of LFR benchmark graphs, each with 5000 vertices and average degree 8.  
The community sizes were selected from the power-law
distribution $f(k) \sim k^{-1.5}$, with $k \in [10,105]$. 
The degree distribution was 
$f(k) \sim k^{-2}$, with $k \in [2,50]$.  
We specified an average degree of 8, which is roughly
comparable to that of the WWW.

Figure ~\ref{fig:mutual} 
contains the mutual information plot for our
experiments with LFR.  
Our metric for comparison is the Jaccard index~\cite{HanKamber-2006}:
\[J(A,B) = \frac{|A \cap B|}{|A \cup B|}\]
where $A$ is the set of intra-community edges in the LFR ground truth, and
$B$ is the set of intra-community edges in an algorithm solution.
As predicted by the resolution limit argument, CNM, an unweighted modularity
maximization algorithm, is not able to resolve most of the natural communities.
However, even with these more realistic data, wCNM achieves greater accuracy 
than the sophisticated HQcut algorithm.  This is notable, considering the
reputation for poor accuracy recently associated with agglomerative algorithms such
as CNM and its variants~\cite{rz2007}.  The accuracy of our CNM variant, on the 
other hand, is competitive.

We observe for these data that iterating the neighborhood coherence 
weighting provides diminishing marginal returns. However, as we show
below, such iteration does add value.

In addition to the mutual information, we wish to compare 
the distributions of the sizes of communities discovered by CNM and its 
weighted variants to the original distributions used in LFR generation.
It is a challenge to fit empirical
data to heavy-tailed power-law distributions.  However, the discrete
power-law distribution
of community sizes used by LFR is not heavy-tailed.
LFR uses the following precise sampling process to determine ground truth
community sizes:
\begin{enumerate}
\item Compute $k^{-\tau}$, the probability that a community will
	have size $k$. 
\item For all $a \le k \le b$, where $a$ and $b$ bound the community sizes,
	compute the empirical cumulative distribution function for $k$:
	$p_k = \sum_{k'=a}^k {k'}^{-\tau}$.
\item For a uniform random variate $x \in [0,1]$, find the 
	minimum $k'$ such that $p_{k'} \ge x$.
\end{enumerate}
This process continues until the sum of the community sizes exceeds the 
number of vertices, and the final community is truncated.

We approach the problem of testing goodness-of-fit of sets of 
algorithm-generated community sizes by generating visualizations and
performing hypothesis tests.  In both cases, we compare the 
empirical distributions of
community sizes with the untruncated discrete power-law distribution that
underlies the LFR distribution.

For visualization, we generate quantile-quantile plots using the R language~\cite{R} and its {\em quantile()}
function with interpolation type 8.  This is the recommendation of 
Hyndman and Fan~\cite{hf1996}.  Figure~\ref{fig:cdf} shows three such plots:
one LFR instance each of $\mu$ values 0.1, 0.3, and 0.5.  With the moderate
community coherence of $\mu = 0.3$, the wCNM variants track the target 
distribution closely, show a drastic improvement over CNM, and appear to
dominate HQcut.  This latter claim is corroborated by the hypothesis tests
described below.  At $\mu = 0.5$, the advantage over CNM is still clear,
but neither wCNM nor HQcut track the target distribution closely.

To augment our results with statistical evidence, we use the classical
Kolmogorov-Smirnov (K--S) test as described, for example, in~\cite{ks-stat}. 
Our null hypothesis is that the 
algorithm-generated community sizes follow a discrete power-law 
with $\tau = 1.5$.  We computed critical values for each sample size between
10 and 290.  The former sometimes occurs in CNM output because of the 
resolution limit, and the latter sometimes occurs in HQcut output as its
stopping criterion encourages splitting communities with high modularity.
The average number of target communities in our LFR instances is 
roughly 150.  
For each sample size, the critical value is the 95th percentile of computed 
K--S statistic values.  We used 100,000 trials per sample size.

After computing critical values, we evaluated the K--S statistic for each of
our trials at each value of $\mu$.  If we reject the null hypothesis then
we have 95\% confidence 
 that the algorithm results do not follow the discrete power-law distribution, 
Table~\ref{tab:ks} summarizes our results for all instances, broken down by
algorithm type and $\mu$ value.

Both Figure~\ref{fig:cdf} and Table~\ref{tab:ks} expose a phenomenon we
call {\em fracturing}.  We refer to the communities defined by LFR as 
{\em target communities}.  There is no guarantee that target communities
will be minimal natural communities.  In fact, the subgraph induced by 
a target community is itself a random graph, and therefore we expect these
to contain minimal natural communities occasionally.  Modularity-based
algorithms such as CNM, wCNM, and HQcut will find these smaller communities
when they exist.  In Table~\ref{tab:ks}, note that wCNM$_5$ fails more K--S
tests than does wCNM$_3$ with increasing $\mu$.  As we add more iterations 
to the
edge weighting scheme described in Section~\ref{sec:edge-weighting}, we
enable wCNM to resolve smaller communities.  The most plausible explanation
for the increased K--S failure rate of wCNM$_5$, holding $\mu$ constant, is
that we detect smaller communities whose sizes were not drawn from the LFR
power-law.  

Figure~\ref{fig:cdf} (b) corroborates this observation.  Note that for
wCNM$_5$, the 
quantile of target community size 10 corresponds to that of discovered
community size less than 5.  Figure~\ref{fig:cdf} (c) shows that HQcut 
also finds communities smaller than size 10.

Algorithms such as wCNM and HQcut ascribe hierarchical community structure
to a graph based on modularity.  Some members of a large collection of 
random graphs, such as the LFR target communities, will have statistically
significant sub-communities.
Lang~\cite{lang09} uses an information theoretic metric to 
distinguish random graphs from those with community substructure.  We
conjecture that Lang's method will judge some LFR target communities to
be non-random.  Modularity-based algorithms will find substructure in
these cases.

\begin{table}[htb]
\begin{tabular}{|c|c|c|c|c|c|} \hline
          & \multicolumn{5}{|c|}{LFR $\mu$} \\ \hline
Algorithm & 0.1 & 0.2 & 0.3 & 0.4 & 0.5 \\ \hline
CNM &0/29 &0/30 &0/30 &0/30 &0/29 \\ \hline
wCNM\_1 &17/29 &0/30 &0/30 &0/30 &0/29 \\ \hline
wCNM\_3 &28/29 &29/30 &29/30 &23/30 &0/29 \\ \hline
wCNM\_5 &28/29 &30/30 &14/30 &0/30 &0/29 \\ \hline
HQcut &12/29 &5/30 &2/30 &2/30 &0/29 \\ \hline
\end{tabular}
\caption{\label{tab:ks} This table shows Kolmogorov-Smirnov (K--S) results for
experiments with 5000-vertex LFR instances (\#passed tests/\#instances). The critical
values for the test were derived empirically by computing the K--S statistic
for 100,000 samples, for each possible sample size between 10 and 290 
communities.  
The hypothesis test results presented are at the 95\% 
confidence level.}
\end{table}

We have not included formal running-time comparisons since Ruan and Zhang's 
publically available HQcut implementation is in Matlab 
and our implementation of wCNM is in C/C++.  
For anecdotal purposes, the wCNM runs on our 5000-vertex LFR instances 
took roughly 10s on a 3Ghz workstation,
even with several iterations of weighting.  The HQcut instances took 
5-10 minutes on the same machine, though there were instances 
that took many hours.  We killed such instances, and that is why 
we sometimes present fewer than 30 instances of HQcut results per $\mu$.

\section{Conclusions}

We agree with Arenas, Fernandez, and Gomez~\cite{afg2008} that it may be 
premature to dismiss the idea of modularity
maximization as a technique for detecting small communities in large networks.
Our weighted analogue to Fortunato and Barth\`{e}lemy's resolution argument
leaves open the possibility for much greater community resolution, given
proper weighting.
Furthermore, our simple adaptation of the CNM heuristic, when combined with a 
careful computation
of edge weights, is able to resolve communities of varying sizes in test 
data.  Furthermore, we have given empirical evidence that the true 
ability of such techniques to resolve small, local communities may be 
greater than that suggested by analysis.

Arguably, the original, unweighted CNM already provides output that could help 
mitigate the resolution limit.
This agglomerative heuristic constructs a dendrogram of hierarchical
communities, and therefore does recognize small communities as modules
before merging them into larger communities.  
In this sense, these small
communities actually are ``resolved'' -- they are stored in the dendrogram
included in the CNM output.  A cut through this 
dendrogram defines the community assignments.
The resolution limit leads us to expect that the communities 
defined by this cut will be unnaturally large.  One potential research 
direction would be to mine this dendrogram for the true communities.
In effect, this would mean ignoring the cut provided by CNM, and therefore
abandoning the idea of maximizing modularity.

Our wCNM heuristic likewise produces a dendrogram and a cut through that 
dendrogram defining communities.  However, the cut provided by wCNM is 
much deeper and more uneven.  It is analagous to the potential result of
mining the CNM dendrogram for natural communities, yet the tie with 
modularity is maintained since wCNM's solution exhibits a maximal 
weighted modularity.

The edge weighting we describe is just one of many possible alternatives,
and wCNM is just one of many potential weighted modularity algorithms.
The main contribution of this paper is to spread awareness that resolution
limits may in fact be tolerated while retaining the advantages of 
modularity maximization and the efficient algorithms for this computation.

\begin{acknowledgments}
We thank Santo Fortunato (ISI), Joseph McCloskey (DoD), 
Cris Moore (UNM), Tamara Kolda (Sandia), 
Dan Nordman (Iowa St.), and Alyson Wilson (Iowa St.)
for helpful discussions and comments.
Sandia is a multiprogram laboratory operated by Sandia Corporation, a Lockheed
Martin Company, for the United States Department of Energy's National 
Nuclear Security Administration under contract DE-AC04-94Al85000.
This work was funded under the Laboratory Directed Research and 
Development program.
\end{acknowledgments}
\bibliography{pre_v2}

\end{document}